\documentclass[pre,twocolumn,showkeys,showpacs,superscriptaddress]{revtex4-1}
\usepackage{graphicx,latexsym,amssymb,amsmath,amsbsy}

\begin{document}

\title{Pressure Distribution and Critical Exponent in Statically Jammed and Shear-Driven Frictionless Disks}

\author{Daniel V{\aa}gberg}
\affiliation{Department of Physics, Ume{\aa} University, 901 87 Ume{\aa}, Sweden}
\author{Yegang Wu}
\affiliation{Department of Physics and Astronomy, University of Rochester, Rochester, NY 14627}
\author{Peter Olsson}
\affiliation{Department of Physics, Ume{\aa} University, 901 87 Ume{\aa}, Sweden}
\author{S. Teitel}
\affiliation{Department of Physics and Astronomy, University of Rochester, Rochester, NY 14627}
\date{\today}

\begin{abstract}
We numerically study the distributions of global pressure that are found in ensembles of statically jammed and quasistatically sheared systems of bidisperse, frictionless, disks at fixed packing fraction $\phi$ in two dimensions.  We use these distributions to address the question of how pressure increases as  $\phi$ increases above the jamming point $\phi_J$, $p\sim |\phi - \phi_J|^y$.  For statically jammed ensembles, our results are consistent with the exponent $y$ being simply related to the power law of the interparticle soft-core interaction.  For sheared systems, however, the value of $y$ is consistent with a non-trivial value, as found previously in rheological simulations.
\end{abstract}
\pacs{45.70.-n, 64.70.Q-, 64.60.-i}
\maketitle

\section{Introduction}

When particles interacting with a short-range repulsive contact potential are confined within a box, a sharp jamming transition takes place as the density of particles is increased \cite{jam}.  Defining the packing fraction as
\begin{equation}
\phi = N\bar v/V
\end{equation}
where $N$ is the total number of particles, $\bar v$ is the average volume per particle, and $V$ is the total system volume, the jamming transition takes place at a critical value $\phi_J$.  For $\phi<\phi_J$, particles pack with no overlaps and the total potential energy $E$ of the system vanishes.  For $\phi>\phi_J$, soft-core particles have some degree of overlap and the resulting contact forces cause $E$ to become finite.  When such a configuration relaxes to a mechanically stable state, the sum of contact forces on each particle vanishes, and the system is at a local minimum of the total potential energy.  Such a configuration is said to be statically jammed.

For an ensemble of such statically jammed {\it frictionless} spheres in two and three dimensions, with average {\it isotropic} stress,
it was observed numerically \cite{OHern} that the pressure $p$ of the total system increased continuously from zero as a power law, as the packing fraction $\phi$ increased above the critical jamming fraction $\phi_J$, 
\begin{equation}
p\sim |\phi - \phi_J|^y\enspace.
\label{e1}
\end{equation}
It was found  by O'Hern {\it et al.} \cite{OHern} that the power law exponent $y$ was simply related to the form of the soft-core contact interaction between overlapping particles.  For a contact potential between two particles,
\begin{equation}
V(r_{ij})=\left\{
\begin{array}{lll}
\dfrac{\epsilon}{\alpha}\left(1-\dfrac{r_{ij}}{d_{ij}}\right)^\alpha\enspace& {\rm for}& r_{ij}<d_{ij}\\\\
0\enspace& {\rm for}& r_{ij}>d_{ij}
\end{array}
\right.
\label{eV}
\end{equation}
it was found that 
\begin{equation}
y=\alpha -1\enspace.
\label{ey}
\end{equation}
Here $r_{ij}=|{\bf r}_{ij}|\equiv |{\bf r}_i-{\bf r}_j|$ is the center-to-center distance between the two particles $i$ and $j$, $d_{ij}=R_i+R_j$ is the sum of the radii of the two particles, and $\epsilon$ is a coupling constant that sets the energy scale.  For a simple harmonic repulsion with $\alpha=2$, we therefore have $y=1$.

If the same system is sheared at a uniform constant shear strain rate $\dot\gamma$, a non-zero pressure $p(\phi,\dot\gamma)$ results for any $\phi$.  The shear-driven jamming $\phi_J$ can be defined \cite{Vagberg} by taking the limit $\dot\gamma\to 0$, where one finds,
\begin{equation}
\lim_{\dot\gamma\to 0}p(\phi,\dot\gamma)= \left\{
\begin{array}{lll}
0& {\rm for} &\phi<\phi_J\enspace,\\\\
p_0(\phi)& {\rm for} &\phi>\phi_J\enspace.  
\end{array}
\right.
\end{equation}
Here  $p_0(\phi)$ is the finite pressure along the yield stress curve separating statically jammed states ($p<p_0$) from states in steady state shear flow ($p>p_0$).  This $p_0(\phi)$ is found to obey a similar power law behavior,
\begin{equation}
p_0(\phi)\sim|\phi-\phi_J|^{y}\enspace.
\end{equation}

It had generally been assumed that the pressure $p_0$ along the yield stress curve, and the pressure $p$ within statically jammed states, should behave similarly, in particular that the exponent $y$ is the same for both cases.  However simulations \cite{OlssonTeitel} of the shearing rheology of overdamped frictionless disks in two dimensions with a harmonic interaction $\alpha=2$ found $y\approx 1.1$, greater than the expected value of Eq.~(\ref{ey}) for statically jammed states, $\alpha-1=1$.  

Since this shearing value of $ y >\alpha-1$ resulted from a detailed critical scaling analysis, which was more complicated than usual because of the need to include correction to scaling terms, it is useful to see if one can find an independent, simpler, analysis that can confirm this result.  In this work we argue that the conclusion, whether or not $y>1$, may be easily obtained by looking at the histogram of total system pressure over the  ensemble of configurations at fixed packing fraction $\phi$.  We first present our numerical results, then we present a simple model to explain them.

\section{Numerical Results}

Our model is one that has been well studied previously in the literature \cite{OHern}. We use a bidisperse mixture of frictionless circular disks in two dimensions (2D), with equal numbers of big and small particles and diameter ratio $d_b/d_s=1.4$.  Particles interact with the soft-core contact potential of Eq.~(\ref{eV}).  We use a fixed number of particles $N$ in a square box with side length $L$, with periodic boundary conditions.  $L$ is adjusted to vary $\phi$.
We will measure length in units such that $d_s=1$, and energy in units such that $\epsilon=1$. 

We consider two different ensembles.  In the first, which we denote as RAND, particles at a fixed $\phi$ are placed at random initial positions, and then quenched to a local energy minimum using a conjugate gradient method.  If the energy per particle of the resulting configuration is $E/N>10^{-16}$ we consider the configuration to be jammed.  Our RAND ensemble is formed by the energy quenched jammed configurations obtained from a large number of independent random initial configurations.  Depending on system size, value of $\phi$, and type of soft-core interaction, our histograms represent between $5000-20000$ independent samples. This is the ensemble considered originally by O'Hern {\it et al.} \cite{OHern}.  Configurations obtained this way, in a fixed square box with periodic boundary conditions, may contain some small residual shear stress.  However on average the stress tensor is one of isotropic pressure.  Such states therefore model the statically jammed states that lie below the yield stress curve.

The second ensemble, which we denote as QS, is obtained by quasistatically shearing \cite{Vagberg,Heussinger,Heussinger2} the system at fixed $\phi$.  Starting from an initial random configuration, we apply an affine finite shear strain $\Delta\gamma$ using Lees-Edwards boundary conditions \cite{LeesEdwards}.  Following the strain step we then use a conjugate gradient  method to relax the strained system to its nearest local energy minimum, giving the non-affine response to the strain step.  Repeating the strain and relaxation steps, our ensemble is formed by the energy minimized configurations at the end of each relaxation step.  We have found \cite{Vagberg} that this ensemble of configurations becomes independent of the initial starting configuration, provided one strains to a sufficiently large total shear strain $\gamma$.  Here we use a strain step $\Delta\gamma=10^{-4}$, sufficiently small that our results are independent of $\Delta\gamma$, and we discard an initial $10000$ steps, corresponding to a strain of $\gamma=1$, to allow the system to reach steady state.  Depending on system size and value of $\phi$ our histograms represent systems sheared to a total strain of roughly $\gamma=10-20$, averaging over $10-30$ independent starting configurations.
The QS ensemble represents states along the yield stress curve $\dot\gamma\to 0$.   Further details of our numerical procedure may be found in Ref.~\cite{Vagberg} (see Sec. II and the Appendix).  

In both ensembles we compute the total system pressure $p$ of each configuration in the usual way \cite{OHern} from the trace of the stress tensor ${\bf P}$ given by the contact forces ${\bf F}_{ij}=-(dV/dr_{ij}){\bf\hat r}_{ij}$,
\begin{equation}
{\bf P}\equiv\dfrac{1}{L^2}\sum_{i,j}{\bf r}_{ij}{\bf F}_{ij}\enspace,\quad p = \frac{1}{2}{\rm Tr}[{\bf P}]\enspace.
\end{equation}

\subsection{RAND Ensemble}

\begin{figure}[h!]
\begin{center}
\includegraphics[width=3.5in]{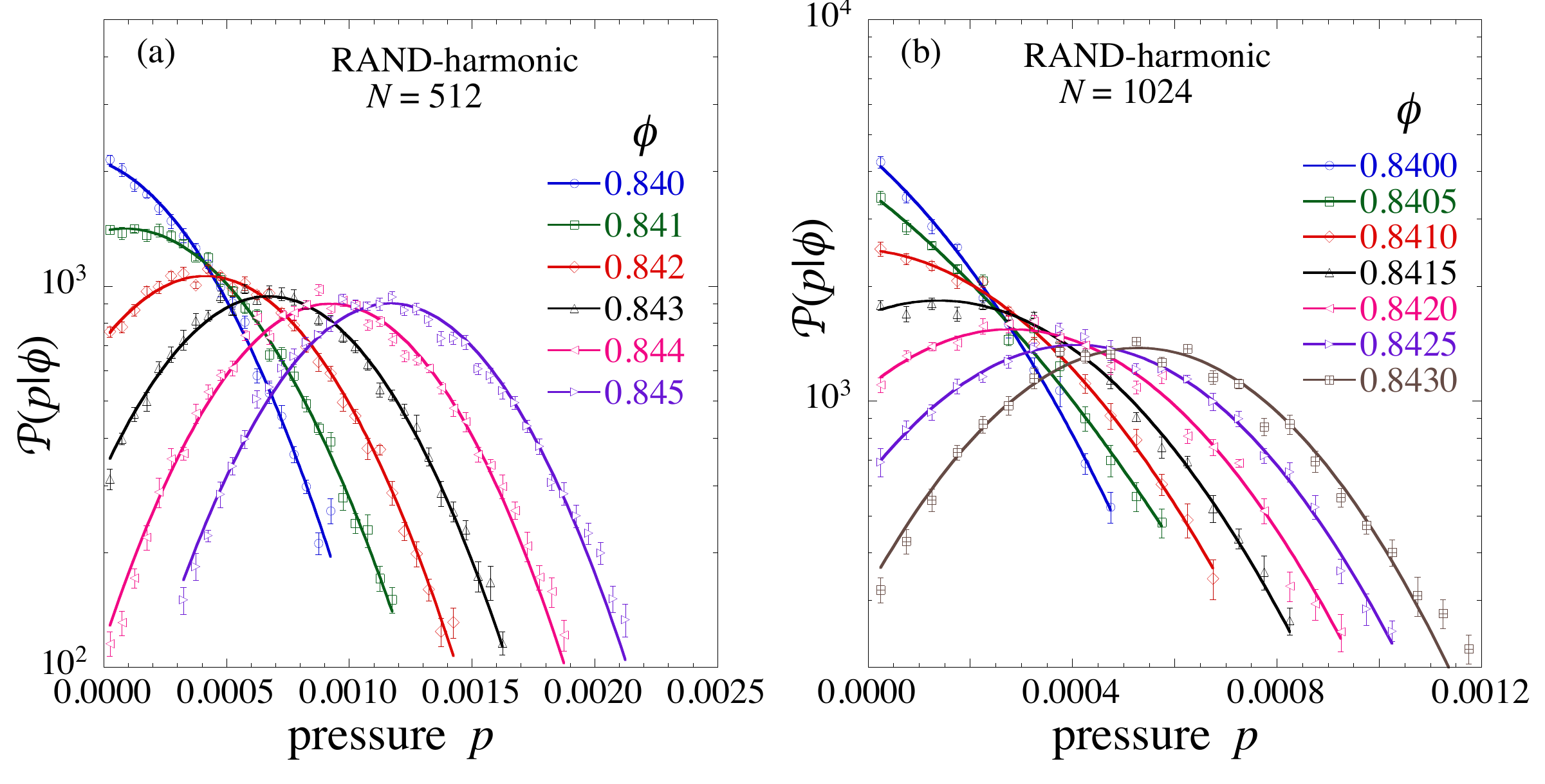}
\caption{Histograms of total system pressure $p$ for several different packing fractions $\phi$, for the RAND ensemble with harmonic soft-core interactions.  The number of particles is: (a) $N=512$, (b) $N=1024$.  Solid lines represent fits to Eq.~(\ref{ePp}) holding $y=1$ fixed.
}
\label{f1}
\end{center}
\end{figure}

In Fig.~\ref{f1} below we show the resulting histograms of pressure, ${\cal P}(p|\phi)$, found for the RAND ensemble at packing fraction $\phi$, with harmonic soft-core potential $\alpha=2$.  In this case we expect from Eq.~(\ref{ey}) that the pressure exponent is $y=1$.  We show results for systems with $N=512$ and $N=1024$ particles, for several different packing fractions $\phi$ close to the value $\phi_J=0.842$, the limiting $N\to\infty$ value of the jamming packing fraction for RAND in 2D \cite{Vagberg2}.  Only jammed configurations with a finite $p>0$ are included in the histograms.  Because we have a finite size system, such jammed configurations exist both below as well as above the $N\to\infty$ value of $\phi_J$.

\begin{figure}[h!]
\begin{center}
\includegraphics[width=3.5in]{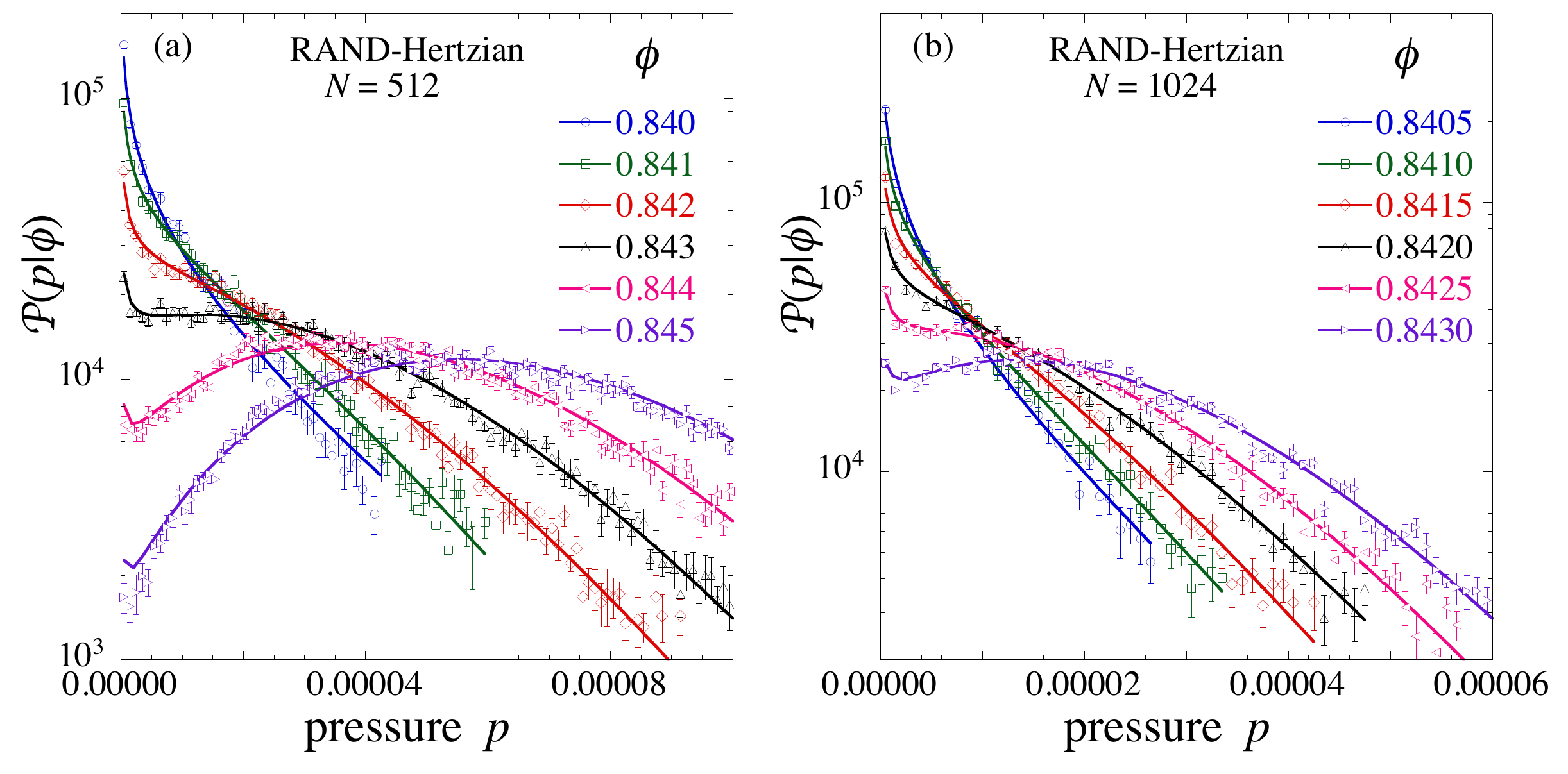}
\caption{Histograms of total system pressure $p$ for several different packing fractions $\phi$, for the RAND ensemble with Hertzian soft-core interactions.  The number of particles is: (a) $N=512$, (b) $N=1024$. Solid lines represent fits to Eq.~(\ref{ePp}) holding $y=3/2$ fixed.
}
\label{f2}
\end{center}
\end{figure}

In Fig.~\ref{f2} we show histograms ${\cal P}(p|\phi)$ for the RAND ensemble, but now for a Hertzian soft-core contact potential, with $\alpha=5/2$.  For this case we expect from Eq.~(\ref{ey}) that $y=\alpha-1=3/2$.  Comparing Figs.~\ref{f1} and \ref{f2} we see a clear qualitative difference.  Whereas for the harmonic interaction ${\cal P}(p|\phi)$ appears to behave smoothly as $p\to 0$, for the Hertzian interaction we see a clear upturn of ${\cal P}(p|\phi)$ as $p$ decreases to small values, suggesting a divergence as $p\to 0$.

\subsection{QS Ensemble}

\begin{figure}[h!]
\begin{center}
\includegraphics[width=3.5in]{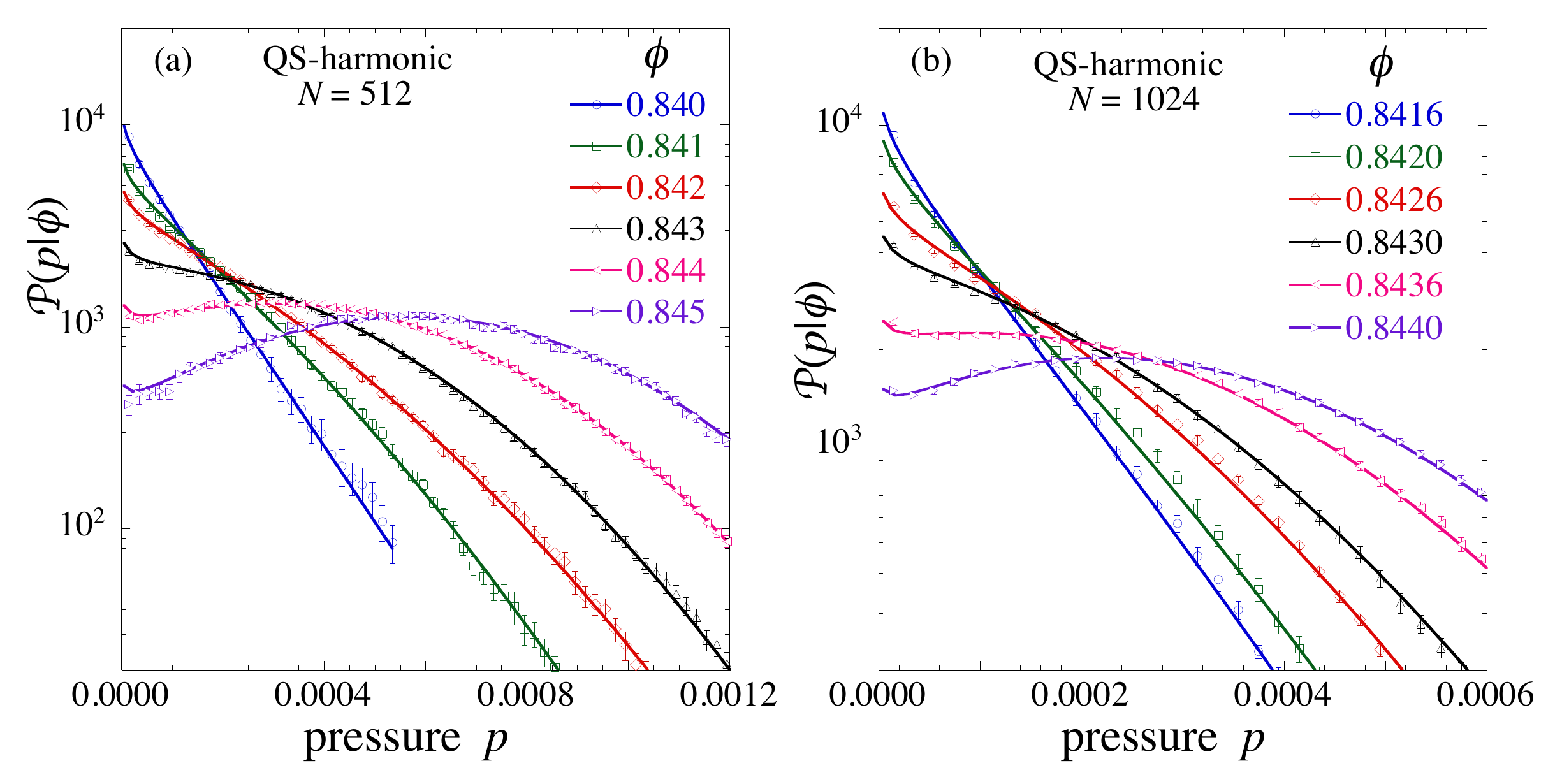}
\caption{Histograms of total system pressure $p$ for several different packing fractions $\phi$, for the QS ensemble (quasistatic shearing) with harmonic soft-core interactions.  The number of particles is: (a) $N=512$, (b) $N=1024$. Solid lines represent fits to Eq.~(\ref{ePp}) holding $y=1.1$ fixed.
}
\label{f3}
\end{center}
\end{figure}

In Fig.~\ref{f3} we show histograms ${\cal P}(p|\phi)$ for the QS ensemble, for the harmonic interaction $\alpha=2$.  We show results for several different packing fractions $\phi$ close to the value $\phi_J=0.843$, the liming $N\to\infty$ value of the {\it shear-driven} jamming transition in 2D \cite{OlssonTeitel,Vagberg2}.  Only configurations with a finite $p>0$ are included in the histograms.  For the harmonic interaction, Eq.~(\ref{ey})
would lead us to expect a value of $y=\alpha-1=1$; however our earlier detailed critical scaling analysis of shearing rheology \cite{OlssonTeitel} resulted in the value $y\approx 1.1$.  Looking at the histograms in Fig.~\ref{f3} we see a clear upturn in ${\cal P}(p|\phi)$ as $p$ decreases to small values, suggesting a possible divergence as $p\to 0$, just as was seen in Fig.~\ref{f2} for the RAND-Hertzian case where $y>1$.  We have verified that this behavior is not an artifact of the bin size chosen to construct the histogram. We may therefore conjecture that the divergence of ${\cal P}(p|\phi)$ as $p\to 0$ is a signature of a pressure exponent $y>1$.

\section{Model}

In this section we propose a simple model to explain the connection between the pressure exponent $y$ and the small $p$ behavior of the pressure histograms ${\cal P}(p|\phi)$.
Consider a statically jammed configuration $i$ in the RAND ensemble, under isotropic stress at a fixed packing fraction $\phi$.  If the box containing the system is slowly and uniformly expanded so as to decrease $\phi$, one will find that the energy decreases and vanishes at some {\it configuration specific} unjamming fraction $\phi_{Ji}$.  
For a system with a finite number of particles $N$, at a fixed initial $\phi$, the values of these $\phi_{Ji}$ form a distribution with finite width as one varies over the configurations $i$ of the ensemble.  The width of the distribution vanishes only as $N\to \infty$.  We denote this distribution as ${\cal P}_J(\phi_{Ji}|\phi)$, the probability that a jammed 
configuration at packing fraction $\phi$ will unjam 
at the packing fraction $\phi_{Ji}$.  Next we will assume that pressure $p$ in such a configuration $i$ is determined by its distance from $\phi_{Ji}$,
\begin{equation}
p=g(\phi-\phi_{Ji})\enspace, \quad{\rm with}\quad g(0)=0\enspace.
\label{eg}
\end{equation}
As found numerically by O'Hern {\it et al.} \cite{OHern} and by Chaudhuri {\it et al.} \cite{Chaudhuri}, we will assume that for $N$ sufficiently large the function $g(\cdot)$ is approximately the same for all configurations $i$.  

We can imagine a similar scenario for a configuration $i$ in the QS ensemble, at its configuration specific yield stress at packing fraction $\phi$.  We can in principle slowly increase the size and perturb the skew of the box so as to decrease the packing fraction $\phi$ while remaining at the configuration specific yield stress; the yield stress should then vanish at a configuration specific $\phi_{Ji}$. The values of $\phi_{Ji}$ obtained this way then give a distribution ${\cal P}_J(\phi_{Ji}|\phi)$, and the pressure $p_0(\phi)$ along the configuration specific yield curve is given by a $g(\phi-\phi_{Ji})$, similar to what was assumed above for RAND.

With this framework in mind, we can then invert Eq.~(\ref{eg}) to write,
\begin{equation}
\phi_{Ji}(p)=\phi- g^{-1}(p)\enspace.
\label{egi}
\end{equation}
It then follows that the probability that a configuration $i$ at packing fraction $\phi$ will be found to have a pressure $p$ is just,
\begin{equation}
{\cal P}(p|\phi)={\cal P}_J(\phi_{Ji}(p)|\phi)\left|\dfrac{d\phi_{Ji}}{dp}\right|\enspace.
\end{equation}
Next we assume, as in Eq.~(\ref{e1}), that
\begin{equation}
g(x)\sim x^y\enspace, \>\>{\rm as}\>\>x\to 0\enspace, \>\>{\rm so\>that}\>\> g^{-1}(p)\sim p^{1/y}\enspace.
\label{egx}
\end{equation}
We then have
\begin{equation}
\left|\dfrac{d\phi_{Ji}}{dp}\right|=\left|\dfrac{-dg^{-1}(p)}{dp}\right|\sim p^{-(1-1/y)}\enspace,
\end{equation}
and so as $p\to 0$,
\begin{equation}
\lim_{p\to 0}{\cal P}(p|\phi)\sim {\cal P}_J(\phi_{Ji}(p)|\phi)p^{-(1-1/y)}\sim p^{-(1-1/y)}\enspace.
\label{ep}
\end{equation}
In the last step we have used that ${\cal P}_J(\phi_{Ji}(p=0)|\phi)={\cal P}_J(\phi|\phi)$ is finite \cite{note2}.

Thus, for $y=1$ we expect ${\cal P}(p|\phi)$ to be finite as $p\to 0$, but for
$y>1$ the pressure distribution diverges algebraically as $p\to 0$.  The presence or absence of such a divergence in ${\cal P}(p|\phi)$ at small $p$ is thus a simple test of whether $y>1$ or $y\le 1$.  This conclusion is in complete agreement with the behavior of ${\cal P}(p|\phi)$ observed in Figs.~\ref{f1}-\ref{f3} if we accept the previously determined values of $y$ found for the three different ensembles.

Note, the direct numerical determination of the function $g(x)$ for individual configurations is somewhat problematic.  For statically jammed configurations under isotropic stress, as in RAND, slowly varying the packing fraction can on occasion trigger an instability that causes a sudden rearrangement of many particles with an accompanying discontinuous jump in pressure.  This tends to be more of a problem upon compressing rather than decompressing \cite{Chaudhuri}.  However for decompressing a configuration along its yield stress curve, as in QS, the difficulty is greatly increased. Firstly, the location of the yield stress curve is not apriori known, and so the trajectory in the $(\phi,\sigma)$ plane (with $\sigma$ the shear stress) that one is trying to follow must be determined in some self-consistent way.  But more importantly, a configuration at its yield stress is inherently at the cusp of going unstable.  If during decompression the system parameters are varied in a way that contains any overlap with the unstable direction in phase space, the particles will suffer large rearrangements and the pressure will jump discontinuously.  In practice we have found that it is possible to continuously (i.e. without large particle rearrangements) decompress configurations along the yield stress curve only over small intervals of $\phi$ too restrictive to be able to accurately determine the exponent $y$ assumed in Eq.~(\ref{egx}).  So instead of directly  computing $g(x)$ numerically, we take Eqs.~(\ref{eg}-\ref{ep}) as an implicit way to determine $g(x)$ from the well defined pressure histograms ${\cal P}(p|\phi)$.

\section{Attempted Data Fitting}

We would like to be able to fit the histograms in the above figures, so as to independently determine a numerical value of the exponent $y$ in each of the different cases.  However such an analysis is complicated by several factors: (i) the range of data for which we see the upturn at small $p$, where the distribution is dominated by the small $p$ power law divergence, is exceedingly narrow; (ii) we do not apriori know the form of ${\cal P}_J(\phi_{Ji}|\phi)$ in Eq.~(\ref{ep}) that we need to do a fitting over a wider range of $p$; (iii) we do not apriori know how the function $g^{-1}(p)$ of Eq.~(\ref{egi}) may depart from a pure power law as $p$ increases from zero to larger values.
As we explain below, we find that (i) and (ii) combine to be too severe a problem to allow us to make a meaningful quantitative estimate of $y$ from our histogram data.

The simplest and most natural guess for the probability distribution ${\cal P}_J(\phi_{Ji}|\phi)$ is a Gaussian,
\begin{equation}
{\cal P}_J(\phi_{Ji}|\phi)\propto {\rm e}^{-\frac{1}{2}z^2}\enspace,\quad{\rm with}\quad z\equiv \dfrac{\phi_{Ji}-\mu(\phi)}{w(\phi)}\enspace,
\end{equation}
where we allow that the average $\mu$ and width $w$ may depend on $\phi$.  If we further assume a pure power law form for $g(x)$ over the entire range of interest,
\begin{equation}
g(x)=Kx^y\quad\Rightarrow\quad g^{-1}(p)=(p/K)^{1/y}\enspace,
\end{equation}
we can then write
\begin{equation}
{\cal P}(p|\phi)=Cp^{-(1-1/y)}{\rm e}^{-\frac{1}{2}\left[((\phi-\mu)K^{1/y}-p^{1/y})/wK^{1/y}\right]^2}\enspace.
\label{ePp}
\end{equation}
Within the above Gaussian approximation for ${\cal P}_J(\phi_{Ji}|\phi)$, we see that a value $y>1$ gives not only a divergence at small $p$, but also an asymmetric peak about the maximum for the exponential term in ${\cal P}(p|\phi)$.  
If, however, the true ${\cal P}_J(\phi_{Ji}|\phi)$ was asymmetric about its peak, this would give another source of asymmetry about the peak of ${\cal P}(p|\phi)$.  In that case, using the symmetric Gaussian approximation for ${\cal P}_J(\phi_{Ji}|\phi)$ and fitting our data to Eq.~(\ref{ePp}) would result in inaccurate values of $y$.  We find this is in fact the situation.  

To see this point, we consider the RAND ensemble with harmonic soft-core potential.  Here, earlier work \cite{OHern,Chaudhuri} has established the value $y=1$, and indeed the histograms in Fig.~\ref{f1} show no evidence of any divergence at small $p$, as is consistent with $y=1$.  We therefore fit the data in Fig.~\ref{f1} to Eq.~(\ref{ePp}) fixing $y=1$.  The results are shown as the solid lines.  We see that even though $y=1$, the histogram peaks, particularly at the larger values of $\phi$, show a noticeable asymmetry; data lies systematically above the fitted curve on the high $p$ side of the peak, and below the fitted curve on the low $p$ side of the peak.  If we fit this same data to Eq.~(\ref{ePp}) with $y$ as a free fitting parameter, we find values of $y\ge 1$, varying sensitively with the range of $\phi$ that is considered in the fit, as well as with the range of $p$ that is used in the histogram at each particular $\phi$.  
The fitted value of $y$ thus arises from a competition between the true $y$ that determines the limiting small $p$ behavior, and an effective $y$ that is trying to model an unknown asymmetry about the histogram peak that is not captured by our Gaussian approximation for ${\cal P}_J(\phi_{Ji}|\phi)$. 

Thus, while the small $p$ behavior of the histograms in Fig.~\ref{f1} supports the conclusion $y=1$, and the small $p$ behavior in Figs.~\ref{f2} and \ref{f3} supports $y>1$ in those cases, we cannot determine reliable numerical values for $y$ from our present histogram data.  We can, however, use previously determined values of $y$ and check for consistency.  Fixing $y=1$ \cite{OHern} for the RAND-harmonic data of Fig.~\ref{f1}, $y=3/2$ \cite{OHern} for the RAND-Hertzian data of Fig.~\ref{f2}, and $y=1.1$ \cite{OlssonTeitel} for the QS-harmonic data of Fig.~\ref{f3}, we fit to Eq.~(\ref{ePp}) and show the results as the solid lines in the respective figures.  We see reasonable eyeball agreement.  

\begin{figure}[h!]
\begin{center}
\includegraphics[width=3.5in]{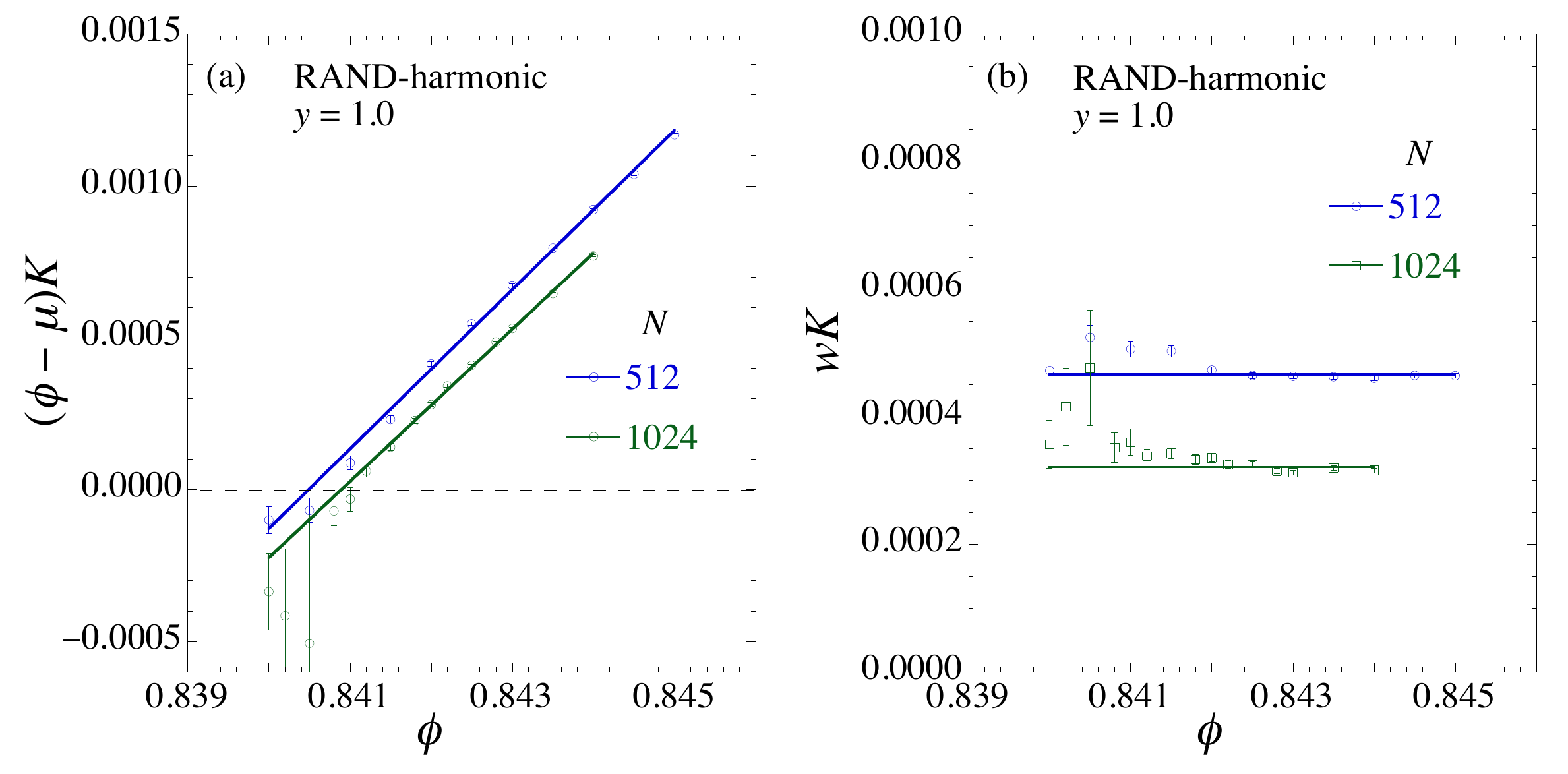}
\caption{Values of (a) $(\phi-\mu)K^{1/y}$ and (b) $wK^{1/y}$, vs $\phi$, for systems with $N=512$ and $1024$ particles, as obtained from fits of the RAND-harmonic data of Fig.~\ref{f1} to Eq.~(\ref{ePp}) keeping $y=1$ fixed.  The solid line in (a) is the best linear fit, while the solid line in (b) is the best fit to a constant.
}
\label{f4}
\end{center}
\end{figure}

\begin{figure}[h!]
\begin{center}
\includegraphics[width=3.5in]{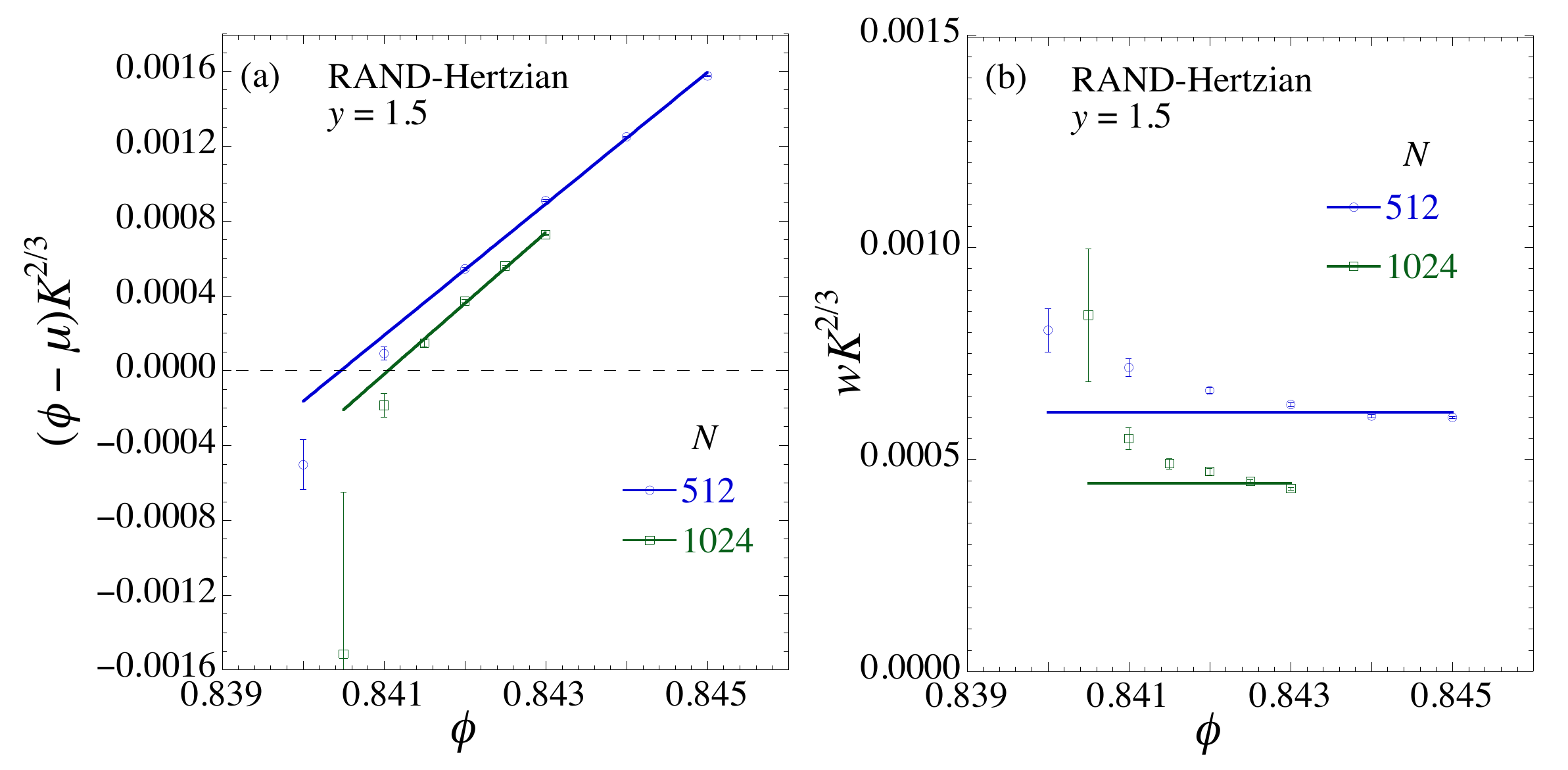}
\caption{Values of (a) $(\phi-\mu)K^{1/y}$ and (b) $wK^{1/y}$, vs $\phi$, for systems with $N=512$ and $1024$ particles, as obtained from fits of the RAND-Hertzian data of Fig.~\ref{f2} to Eq.~(\ref{ePp}) keeping $y=3/2$ fixed. The solid line in (a) is the best linear fit, while the solid line in (b) is the best fit to a constant.
}
\label{f5}
\end{center}
\end{figure} 

\begin{figure}[h!]
\begin{center}
\includegraphics[width=3.5in]{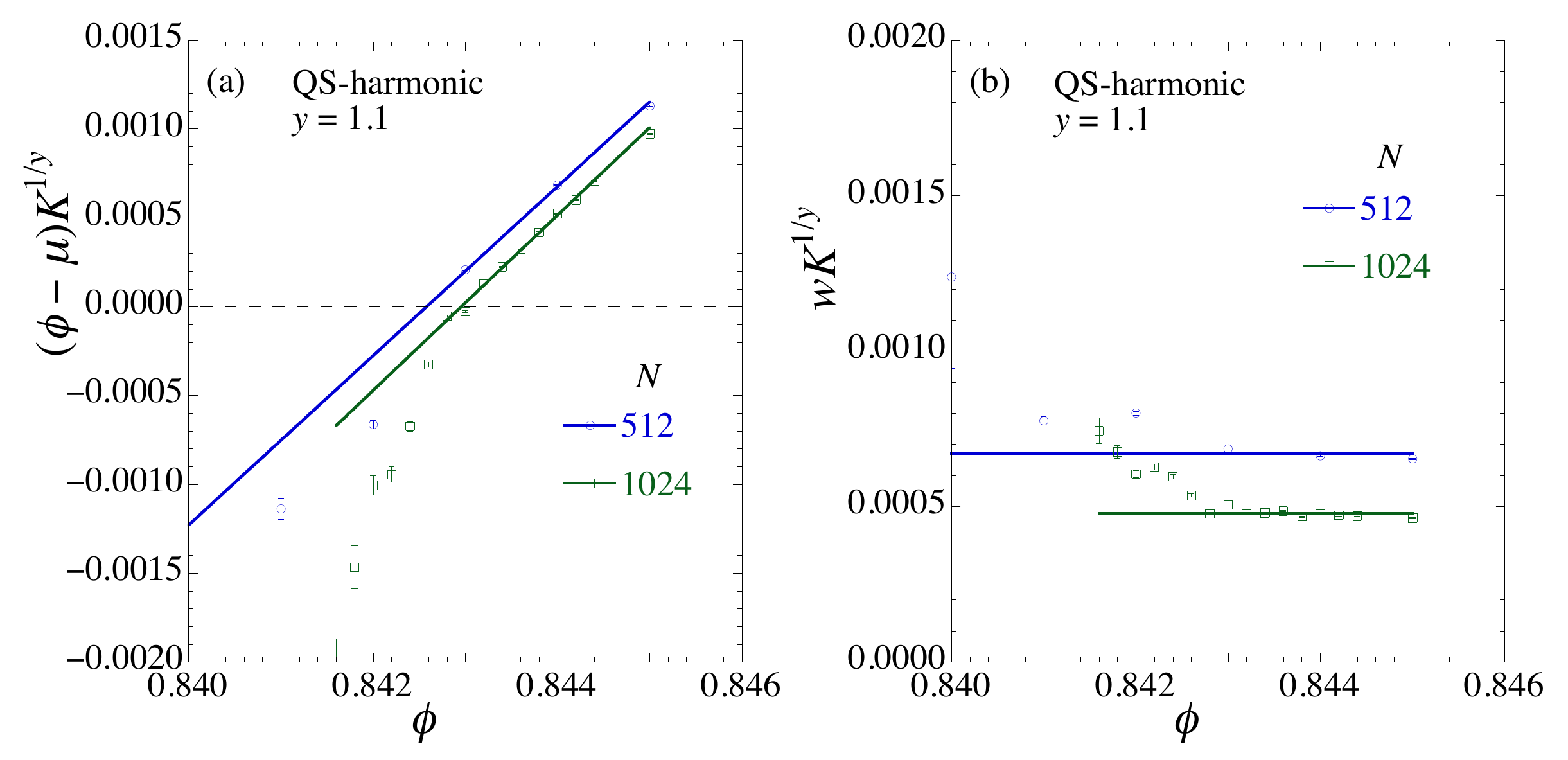}
\caption{Values of (a) $(\phi-\mu)K^{1/y}$ and (b) $wK^{1/y}$, vs $\phi$, for systems with $N=512$ and $1024$ particles, as obtained from fits of the QS-harmonic data of Fig.~\ref{f3} to Eq.~(\ref{ePp}) keeping $y=1.1$ fixed. The solid line in (a) is the best linear fit, while the solid line in (b) is the best fit to a constant.
}
\label{f6}
\end{center}
\end{figure} 

To further test consistency, we plot the values of $(\phi-\mu)K^{1/y}$ and $wK^{1/y}$ obtained from these fits with fixed $y$, versus $\phi$, in Figs.~\ref{f4}, \ref{f5} and \ref{f6}.  In each case we see that $(\phi-\mu(\phi))K^{1/y}$ is roughly linear in $\phi$, at least at the larger values of $\phi$ where the estimated statistical error is  small, indicating a relatively weak dependence of $\mu$ on $\phi$.  Fitting to the form $C(\phi-\phi_J)$, with $C$ and $\phi_J$ as free parameters, we find from Fig.~\ref{f4}(a) for RAND with harmonic interactions: $N=512$, $\phi_J=0.8405$; $N=1024$, $\phi_J=0.8409$.
From  Fig.~\ref{f5}(a) for RAND with Hertzian interactions we find: $N=512$, $\phi_J=0.8405$; $N=1024$, $\phi_J=0.8411$, in reasonable agreement with the harmonic interaction.  As expected, we find the value of $\phi_J$ to be independent of the particular soft-core interaction.  These values are also reasonably consistent with the finite size estimates of $\phi_J$ as obtained from Ref.~\cite{Vagberg2} (see Fig. 1(a))), with $\phi_J$ increasing as $N$ increases.  For QS with harmonic interactions we have from Fig.~\ref{f6}(a): $N=512$, $\phi_J=0.8426$; $N=1024$, $\phi_J=0.8430$.  These values are also reasonably consistent with the finite size estimates of the shear-driven $\phi_J$ from Ref.~\cite{Vagberg2} (see Fig. 1(b)).

Finally we can consider the width parameter $wK^{1/y}$.  We see that in most cases $wK^{1/y}$ is roughly constant at the larger values of $\phi$, and the ratio of widths comparing $N=1024$ to $N=512$ is $0.69$, $0.73$, $0.71$, for RAND-harmonic, RAND-Hertzian, and RAND-QS respectively.  This is in reasonable agreement with the value $1/\sqrt{2}\approx 0.71$ expected for the usual $1/\sqrt{N}$ finite size dependence.

\section{Conclusions}

To conclude, we have presented a simple method, based on ensemble histograms of total system pressure, to determine whether the exponent $y$ with which the system pressure algebraically increases from zero, as $\phi$ increases above $\phi_J$, satisfies $y>1$ or $y\le 1$.  We find results consistent with our earlier finding \cite{OlssonTeitel} that, for harmonically interacting particles, $y>1$ for the pressure along the yield stress curve of shear-driven systems.  This is in contrast to the expectation of Eq.~(\ref{ey}) for statically jammed systems.  While our method is not at present accurate enough to allow a reliable determination of the precise numerical value of $y$, we find our results are consistent with previous determined values.

\section*{Acknowledgements}

This work was supported by NSF grant DMR-1205800 and Swedish Research Council grant 2010-3725. Simulations were performed on resources
provided by the Swedish National Infrastructure for Computing (SNIC) at PDC and HPC2N, and the Center for Integrated Research Computing (CIRC) at the University of Rochester.


\begin{thebibliography}{99}
\bibitem{note2}Note that since a system at its configuration specific jamming fraction $\phi_{Ji}$ has pressure $p=0$, then ${\cal P}_J(\phi_{Ji}(p=0)|\phi) = {\cal P}_J(\phi|\phi)$, i.e. the probability that a system at packing fraction $\phi$ is exactly at its configuration specific jamming point.  The integral of this quantity, $f(\phi)=\int_0^\phi d\phi^\prime {\cal P}_J(\phi^\prime |\phi^\prime)$ is then just the fraction of jammed states at $\phi$, as we have numerically computed earlier, see Fig.1 of Ref.~\cite{Vagberg2}.  Those results confirm that for finite system sizes, ${\cal P}_J(\phi|\phi)$ is always finite.
\end{thebibliography}
\end{document}